\documentstyle[12pt]{article}
\begin{document}

\def\Bbb{\bf }

\title{Coloring the rational quantum sphere and the Kochen-Specker theorem}
\author{Hans Havlicek \\
 {\small Institut f\"ur Geometrie,}
  {\small Technische Universit\"at Wien   }     \\
  {\small Wiedner Hauptstra\ss e 8-10/1133,}
  {\small A-1040 Vienna, Austria   }            \\
  {\small havlicek@geometrie.tuwien.ac.at}   \\
{\small and}\\
G\"unther Krenn and
Johann Summhammer\\
{\small Atominstitut der \"osterreichischen Universit\"aten,}\\
  {\small Stadionallee 2}                  \\
  {\small A-1020 Vienna, Austria   }            \\
  {\small krenn@ati.ac.at, summhammer@ati.ac.at}\\
{\small and}\\
Karl Svozil\\
 {\small Institut f\"ur Theoretische Physik,}
  {\small Technische Universit\"at Wien }     \\
  {\small Wiedner Hauptstra\ss e 8-10/136,}
  {\small A-1040 Vienna, Austria   }            \\
  {\small e-mail: svozil@tuwien.ac.at}\\
  {\small (to whom correspondence should be directed)}}
\date{ }
\maketitle

\begin{flushright}
{\scriptsize http://tph.tuwien.ac.at/$\widetilde{\;\;}\,$svozil/publ/2000-co.$\{$htm,ps,tex$\}$}
\end{flushright}

\begin{abstract}
We review and extend
recent findings of Godsil and Zaks \cite{godsil-zaks}, who
published a constructive coloring
of the rational unit sphere
with the property that for any
orthogonal tripod formed by rays
extending from the origin of the points of the sphere, exactly one ray
is red,  white and black.
They also showed that any consistent coloring of the real sphere
requires an additional color.
We discuss some of the consequences for the Kochen-Specker theorem \cite{kochen1}.
\end{abstract}

\newpage
\section{Colorings}

In what follows we shall consider ``rational rays.''
A ``rational ray'' is  the linear span of a non-zero vector
of ${\Bbb Q}^n \subset {\Bbb R}^n$.

Let $p$ be a prime number. A coloring of the rational rays
of ${\Bbb R}^n$, $n\geq 1$, using $p^{n-1}+p^{n-2}+\ldots+1$
colors can be constructed in a straightforward manner. We
refer to \cite{klingenberg,hales,carter} for the theoretical background of the
following construction.

Each rational ray is the linear span of a vector
$(x_1,x_2,\ldots,x_n)\in{\Bbb Z}^n$, where
$x_1,x_2,\dots,x_n$ are coprime. Such a vector is unique up
to a factor $\pm 1$.

Next, let ${\Bbb Z}_p$ be the field of residue classes
modulo $p$. The vector space ${\Bbb Z}_p^n$ has $p^n -1$
non-zero vectors; each ray through the origin of ${\Bbb
Z}_p^n$ has $p-1$ non-zero vectors. So there are exactly
$(p^n -1)/(p-1)= p^{n-1}+p^{n-2}+\ldots+1$ distinct rays
through the origin which can be colored with
$p^{n-1}+p^{n-2}+\ldots+1$ distinct colors.

Finally, assign to the ray $Sp(x_1,x_2,\ldots,x_n)$
(``$Sp$'' denotes linear span) the color of the ray of
${\Bbb Z}_p^n$ which is obtained by taking the modulus of
the coprime integers $x_1,x_2,\ldots,x_n$ modulo $p$.
Observe that $x_1,x_2,\ldots,x_n$ cannot vanish
simultaneously modulo $p$ and that $\pm(x_1,x_2,\ldots,x_n)$
yield the same color. Obviously, all
$p^{n-1}+p^{n-2}+\ldots+1$ colors are actually used.

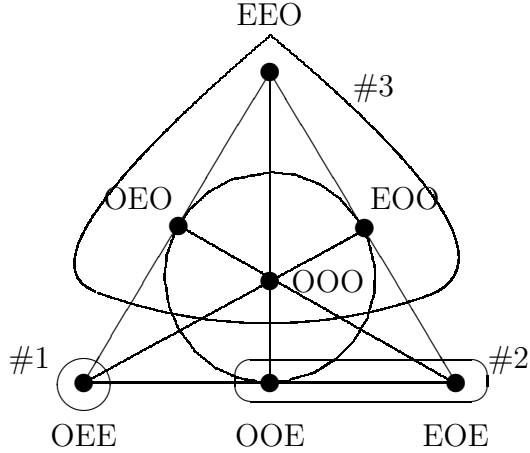
\begin{figure}
\begin{center}
\unitlength 0.700mm
\linethickness{0.4pt}
\begin{picture}(99.67,80.00)
\multiput(50.00,50.00)(0.81,-0.10){6}{\line(1,0){0.81}}
\multiput(54.88,49.40)(0.31,-0.12){15}{\line(1,0){0.31}}
\multiput(59.46,47.62)(0.17,-0.12){24}{\line(1,0){0.17}}
\multiput(63.47,44.78)(0.12,-0.14){27}{\line(0,-1){0.14}}
\multiput(66.67,41.05)(0.12,-0.23){19}{\line(0,-1){0.23}}
\multiput(68.86,36.65)(0.12,-0.53){9}{\line(0,-1){0.53}}
\multiput(69.91,31.85)(-0.08,-2.46){2}{\line(0,-1){2.46}}
\multiput(69.76,26.93)(-0.11,-0.39){12}{\line(0,-1){0.39}}
\multiput(68.42,22.20)(-0.12,-0.20){21}{\line(0,-1){0.20}}
\multiput(65.96,17.95)(-0.12,-0.12){29}{\line(0,-1){0.12}}
\multiput(62.54,14.42)(-0.19,-0.12){22}{\line(-1,0){0.19}}
\multiput(58.36,11.83)(-0.36,-0.11){13}{\line(-1,0){0.36}}
\multiput(53.67,10.34)(-1.64,-0.10){3}{\line(-1,0){1.64}}
\multiput(48.77,10.04)(-0.60,0.11){8}{\line(-1,0){0.60}}
\multiput(43.94,10.94)(-0.25,0.11){18}{\line(-1,0){0.25}}
\multiput(39.47,13.00)(-0.15,0.12){26}{\line(-1,0){0.15}}
\multiput(35.64,16.08)(-0.12,0.16){25}{\line(0,1){0.16}}
\multiput(32.68,20.00)(-0.12,0.28){16}{\line(0,1){0.28}}
\multiput(30.76,24.53)(-0.11,0.69){7}{\line(0,1){0.69}}
\multiput(30.01,29.38)(0.11,1.22){4}{\line(0,1){1.22}}
\multiput(30.46,34.28)(0.12,0.33){14}{\line(0,1){0.33}}
\multiput(32.10,38.91)(0.12,0.18){23}{\line(0,1){0.18}}
\multiput(34.81,43.01)(0.13,0.12){28}{\line(1,0){0.13}}
\multiput(38.44,46.32)(0.22,0.12){20}{\line(1,0){0.22}}
\multiput(42.78,48.65)(0.60,0.11){12}{\line(1,0){0.60}}
\put(50.00,69.00){\line(3,-5){35.40}}
\put(50.00,69.00){\line(-3,-5){35.40}}
\put(14.60,10.00){\line(1,0){70.73}}
\put(50.00,10.00){\line(0,1){59.00}}
\multiput(14.67,10.00)(0.22,0.12){242}{\line(1,0){0.22}}
\multiput(85.00,10.00)(-0.21,0.12){248}{\line(-1,0){0.21}}
\put(32.67,39.67){\circle{2.67}}
\put(32.67,39.67){\circle*{3.33}}
\put(68.00,39.33){\circle*{3.33}}
\put(50.00,29.33){\circle*{3.33}}
\put(50.00,10.00){\circle*{3.33}}
\put(85.33,10.00){\circle*{3.33}}
\put(14.67,10.00){\circle*{3.33}}
\put(50.00,69.00){\circle*{3.33}}
\put(14.67,0.00){\makebox(0,0)[cc]{OEE}}
\put(50.00,0.00){\makebox(0,0)[cc]{OOE}}
\put(85.33,0.00){\makebox(0,0)[cc]{EOE}}
\put(50.00,80.00){\makebox(0,0)[cc]{EEO}}
\put(75.67,45.00){\makebox(0,0)[cc]{EOO}}
\put(25.00,45.00){\makebox(0,0)[cc]{OEO}}
\put(61.00,29.33){\makebox(0,0)[cc]{OOO}}
\bezier{268}(17.33,27.00)(49.33,15.67)(80.33,26.67)
\bezier{344}(80.33,26.67)(99.67,34.67)(50.00,76.00)
\bezier{328}(50.00,76.00)(0.67,33.67)(16.67,27.33)
\put(67.33,10.33){\oval(48.00,8.00)[]}
\put(14.67,9.67){\circle{9.71}}
\put(4.33,14.33){\makebox(0,0)[cc]{\#1}}
\put(95.33,14.33){\makebox(0,0)[cc]{\#2}}
\put(69.67,66.00){\makebox(0,0)[cc]{\#3}}
\end{picture}
\end{center}
\caption{\label{f-coloring1}The projective plane over ${\Bbb Z}_2$.
and the reduced coloring scheme discussed.}
\end{figure}

In what follows, we consider the case $p=2$, $n=3$. Here all
rational rays $Sp(x,y,z)$ (with $x,y,z\in{\Bbb Z}$ coprime)
are colored according to the property which ones of the
components $x,y,z$ are even (E) and odd (O). There are
exactly $7$ of such triples OEE, EOE, EEO, OOE, EOO, OEO,
OOO which are associated with one of seven different colors
$\#1,\#2,\#3,\#4,\#5,\#6,\#7$. Only the EEE triple is
excluded. Those seven colors can be identified with the seven
points of the projective plane over ${\Bbb Z}_2$; cf. Fig. \ref{f-coloring1}.

Next, we restrict our attention to those rays which meet the
rational unit sphere $S^2\cap {\Bbb Q}^3$. The following
statements on a triple $(x,y,z)\in {\Bbb Z}^3 \setminus
\{(0,0,0)\}$ (not necessarily coprime) are equivalent:
\begin{description}
\item{(i)}
The ray $Sp(x,y,z)$ intersects the unit sphere at two
rational points; i.e., it contains the rational points
$\pm\left( x,y,z\right)/\sqrt{x^2+y^2+z^2}\in  S^2\cap {\Bbb
Q}^3$.
\item{(ii)}
The Pythagorean property holds, i.e., $x^2+y^2+z^2 = n^2,\;
n\in {\Bbb N}$.
\end{description}
This equivalence can be demonstrated as follows. All points
on the rational unit sphere can be written as ${\bf
r}=\left({a\over a'},{b\over b'},{c\over c'} \right)$ with
$a,b,c\in {\Bbb Z}$, $a',b',c'\in{\Bbb Z}\setminus \{0\}$,
and $\left({a\over a'}\right)^2+\left({b\over
b'}\right)^2+\left({c\over c'}\right)^2 =1$. Multiplication
of ${\bf r}$ with ${a'}^2{b'}^2{c'}^2$ results in a vector
of ${\Bbb Z}^3$ satisfying (ii). Conversely, from
$x^2+y^2+z^2 = n^2,\; n\in {\Bbb N}$, we obtain the rational
unit vector $\left({x\over n},{y\over n},{z\over n}
\right)\in S^2\cap {\Bbb Q}^3$.

Notice that this Pythagorean property is rather restrictive.
Not all rational rays intersect the rational unit sphere.
For a proof, consider $Sp(1,1,0)$ which intersects the unit
sphere at $\pm (1/\sqrt{2})(1,1,0)\not\in S^2\cap {\Bbb
Q}^3$. Although both the set of rational rays as well as $
S^2\cap {\Bbb Q}^3$ are dense, there are ``many'' rational
rays which do not have the Pythagorean property.

If $x,y,z$ are chosen coprime then a necessary condition for
$x^2+y^2+z^2$ being a non-zero square is that precisely one
of $x$, $y$, and $z$ is odd. This is a direct consequence of
the observation that any square is congruent to 0 or 1,
modulo 4, and from the fact that at least one of $x$, $y$,
and $z$ is odd. Hence our coloring of the rational rays
induces the following coloring of the rational unit sphere
with those three colors that are represented by the standard
basis of ${\Bbb Z}_2^3$:

\begin{description}
\item{color \#1} if $x$ is odd, $y$ and $z$ are even,
\item{color \#2} if $y$ is odd, $z$ and $x$ are even,
\item{color \#3} if $z$ is odd, $x$ and $y$ are even.
\end{description}

All three colors occur, since the vectors of the standard
basis of ${\Bbb R}^3$ are colored differently.

Suppose that two points of $S^2\cap {\Bbb Q}^3$ are on rays
$Sp(x,y,z)$ and $Sp(x',y',z')$, each with coprime entries.
The inner product $xx'+yy'+zz'$ is even if and only if the
inner product of the corresponding basis vectors of ${\Bbb
Z}_2^3$ is zero or, in other words, the points are colored
differently. In particular, three points of $S^2\cap {\Bbb
Q}^3$ with mutually orthogonal position vectors are colored
differently.

From our considerations above, three colors are sufficient
to obtain a coloring of the rational unit sphere
$S^2\cap{\Bbb Q}^3$ such that points with orthogonal
position vectors are colored differently, but clearly this
cannot be accomplished with two colors. So the ``chromatic
number'' for the rational unit sphere is three. This result
is due to Godsil and Zaks \cite{godsil-zaks}; they also
showed that the chromatic number of the real unit sphere is
four. However, they obtained their result in a slightly different way.
Following \cite{hales} all rational rays are associated with three colors
by making the following identification:
\begin{eqnarray}
 \#1, \nonumber \\
 \#2 =\#4, \nonumber \\
 \#3 = \#5 = \#6 = \#7. \nonumber
\end{eqnarray}
This 3-coloring has the property that coplanar rays
are always colored by using only two colors; cf. Fig. \ref{f-coloring1}.
According to our approach this intermediate 3-coloring
is not necessary, since rays in colors
\#4, \#5,  \#6,  \#7 do not meet the rational unit sphere.

\subsection{Reduced two-coloring}
As a corollary, the rational unit sphere can be colored by two
colors such that,
for any arbitrary orthogonal tripod spanned by rays through its origin,
one vector is colored by color \#1
and the other rays are colored
by color \#2.  This can be easily verified by identifying colors \#2 \&
\#3 from the above scheme. (Two equivalent two-coloring schemes
result from a reduced  chromatic three-coloring scheme by requiring that
color \#1 is associated with $x$ or $y$ being odd, respectively.)

Kent \cite{kent:99} has shown that there also exist dense sets in higher dimensions which permit a
reduced two-coloring.
Unpublished results by  P. Ovchinnikov, O.G.Okunev and D. Mushtari
\cite{ovchin-pr:2000}
state
that the rational d-dimensional unit sphere is d-colorable if and only if it admits a
reduced two-coloring if and only if $d<6$.

\subsection{Denseness of single colors}
It can also be shown that each color class in the above coloring schemes
is dense in the sphere. To prove this, Godsil and Zaks consider $\alpha$
such that
$\sin \alpha ={3\over 5 }$ and thus
$\cos \alpha ={4\over 5 }$.
$\alpha$ is not a rational multiple of $\pi$; hence
$\sin (n\alpha)$
and
$\cos (n\alpha)$
are non-zero for all integers $n$. Let $F$ be the rotation matrix about
the $z$-axis through an angle $\alpha$; i.e.,
$$F=
\left(
\begin{array}{ccc}
\cos \alpha&\sin \alpha&0\\
-\sin \alpha&\cos \alpha&0\\
0&0&1\\
\end{array}
\right).
$$
Then the image $I$, under the powers of $F$, of the point $(1,0,0)$ is a
dense subset of the equator.

Now suppose that the point $u=\left( {a\over c},{b\over
c},0\right)$
is on the rational unit sphere and that $a,c$ are odd and thus $b$ is even.
In the coloring scheme introduced above, $u$ has the same color as
$(1,0,0)$ (identify $a=c=1$ and $b=0$);
and so does $Fu$.
This proves that $I$
(the image of all powers of $F$ of the points $u$)
is  dense.
We shall come back to the physical consequences of this property later.

In the reduced two-color setting,
if the two "poles"
$\pm (0,0,1)$
acquire color \#1, then the entire equator acquires color \#2.
Thus, for example, for the two tripods spanned by
$\{(1,0,0),(0,1,0),(0,0,1)\}$ and
$\{(3,4,0),(-4,3,0),(0,0,1)\}$, the first two legs have color \#2, while
$(0,0,1)$ has color \#1.

\subsection{The chromatic number of the real unit sphere in three dimensions is four}

A proof that four colors suffice for the coloring of
points of the unit sphere in three dimensions
is constructive and rather elementary.
Consider first the intersection points of the
sphere with the the $x-$, the $y-$ and the $z-axis$,
colored by green, blue and red, respectively.
There are exactly three great circles which pass through two of these three pairs of points.
The great circles can be colored with the two colors used on the four points they pass through.
The three great circles divide the sphere into eight open octants
of equal area. Four octants, say, in the half space $z>0$, are colored by the four colors
red, white, green and blue. The remaining octants
obtain their color from their antipodal octant.

Although Godsil and Zaks'  \cite{godsil-zaks} paper is not entirely specific,
it is easy to write down an explicit coloring scheme according
to the above prescription.
Consider spherical coordinates: let
$\theta$ be the angle between the $z-$axis and the line connecting the origin and the point, and
$\varphi$ be the angle between the $x-$axis and the projection  of the line connecting the origin and the point
onto the $x-y-$plane.
In terms of these coordinates, an arbitrary point on the unit sphere is given by
$(\theta, \varphi, r=1) \equiv (\theta,\varphi)$.
\begin{itemize}
\item
The colors of the cartesian coordinate axes $(\pi/2,0)$, $(\pi/2,\pi/2)$, $(0,0)$
are
green,
blue and  red, respectively.
\item
The color of the octant
$\{(\theta, \varphi ) \mid 0 < \theta \le \pi /2,\; 0\le \varphi < \pi/2\}$ is green.
\item
The color of the  octant
$\{(\theta, \varphi ) \mid 0 \le \theta < \pi /2,\; \pi /2 \le \varphi \le \pi\}$ is red.
\item
The color of the octant
$\{(\theta, \varphi ) \mid 0 < \theta < \pi /2,\; \pi  < \varphi < 3\pi /2\}$ is white.
\item
The color of the octant
$\{(\theta, \varphi ) \mid 0 < \theta \le \pi /2,\; -\pi /2  \le \varphi < 0\}$ is blue.
\item
The colors of the points in the half space $z<0$ are inherited from their antipodes.
This completes the coloring of the sphere.
\end{itemize}

The fact that three colors are not sufficient is not so obvious.
Here we shall not review Godsil and Zaks' proof based on a paper by
A. W. Hales and E. G. Straus \cite{hales}, but refer to a result of
S. Kochen and E. Specker \cite{kochen1},
which is of great importance in the
present debate on hidden parameters in quantum mechanics.
They have proven that there does not exist a reduced two-coloring, also termed valuation,
on the one dimensional subspaces of real Hilbert space in three dimensions.

Recall that
a reduced two-coloring of the one dimensional linear subspaces
with two colors could immediately be obtained
from any possible appropriate coloring of  the sphere with three colors
by just identifying two of the
three colors. Thus, the impossibility of a reduced two-coloring implies that
three colors are not sufficient for an appropriate coloring of the threedimensional real unit sphere.
(``Appropriate'' here means: ``points at spherical distance $\pi /2$ get different colors.'')

In the same article \cite{kochen1}, Kochen and  Specker
gave an explicit example (their $\Gamma_3$)
of a finite point set of the sphere
with weaker properties which suffice just as well for
this purpose: the structure still allows for a
reduced two-coloring, yet it cannot be colored by three colors.
(The authors did not mention nor discuss this particular feature \cite{Specker-priv}.)

The impossibility  of a reduced two-coloring also rules out another attempt
to ``nullify'' the Kochen-Specker theorem by identifying pairs of colors of an appropriate
four-coloring of the real unit sphere.
Any such identification would result in tripods colored by \#1-\#2-\#2,
as well as for instance  \#1-\#1-\#2,
which is not allowed for reduced coloring schemes, which requires colorings of the type  \#1-\#2-\#2.

\section{Physical aspects}

\subsection{Physical truth values}
Based on Godsil and Zaks´ \cite{godsil-zaks} results, Meyer \cite{meyer:99} suggested that the physical
impact of the Kochen-Specker theorem
\cite{kochen1} is ``nullified,'' since for all practical
purposes it is impossible to operationalize
the difference between any dense set of rays and
the continuum of Hilbert space rays.
(See also the subsequent papers by Kent \cite{kent:99} and Clifton and Kent \cite{clifton:99}.)
However, for the reasons given below, the physical applicability of these constructions remain
questionable.

Let us re-state the physical interpretation of the coloring schemes discussed above.
Any linear subspace $Sp\, {\bf r}$ of a vector ${\bf r}$
can be identified with the associated projection operator $E_{{\bf r}}$
and with the quantum mechanical proposition ``the physical system is in
a pure state $E_{{\bf r}}$'' \cite{v-neumann-49}.
The coloring of the associated point on
the unit sphere (if it exists) is equivalent with a valuation or
two-valued probability measure
$$Pr:E_{{\bf r}}\mapsto \{0,1\}$$
where $0\sim \#2$ and
 $1\sim \#1$.
That is,
the two colors \#1, \#2 can be identified with the
classical truth values:
 ``It is true that the physical system is in
a pure state $E_{{\bf r}}$'' and
 ``It is false that the physical system is in
a pure state $E_{{\bf r}}$,''
respectively.

Since, as has been argued before, the rational unit sphere has
chromatic number three,
two colors suffice for a reduced coloring
generated under the assumption that the colors of two  rays
in any orthogonal tripod are identical.
This effectively generates
consistent valuations associated
with the dense
subset of physical properties corresponding to the rational unit sphere
\cite{meyer:99}.

Kent \cite{kent:99} has shown that there also exist dense sets in higher dimensions which permit a
reduced two-coloring.
Unpublished results by  P. Ovchinnikov, O.G.Okunev and D. Mushtari
\cite{ovchin-pr:2000}
state
that the rational unit sphere of
the d-dimensional real Hilbert space is d-colorable if and only if it admits a
reduced two-coloring if and only if $d<6$.


\subsection{Sufficiency}
The Kochen-Specker theorem deals with the nonembedability of
certain partial algebras---in particular the ones arising
in the context of quantum mechanics---into total Boolean algebras.
$0-1$ colorings serve as an important method of realizing such embeddings:
e.g., if there are a sufficient number of them to separate any two elements
of the partial algebra, then embedability follows \cite{kochen1,pulmannova-91,svozil-tkadlec}.
Kochen and Specker's  original paper \cite{kochen1} contains a much stronger result---the
nonexistence of $0-1$ colorings---than would be needed for nonembedability.
However, the mere existence of some homomorphisms
is {\em a necessary but no sufficient criterion} for embedability.
The Godsil and Zaks  construction merely provides three homomorphisms which are not sufficient
to guarantee embedability.
This has already be pointed out by Clifton and Kent \cite{clifton:99}.

\subsection{Continuity}
The regress to unsharp measurements is rather questionable and would
allow total arbitrariness in the choice of approximation.
That is, due to the density of the coloring, depending on which approximation is chosen,
one predicts different results.
The probabilities resulting
from these {\em truth assignments are noncontinuous and arbitrary.}
This is even more problematic if one realizes that
the each color of the $0-1$ coloring of the rational unit sphere is dense.

\subsection{Closedness}

The rational unit sphere is not closed under certain geometrical operations such
as taking an orthogonal ray of the subspace spanned by two
non-collinear rays (the cross product of the associated vectors).
This can be easily demonstrated by considering the two vectors
$$\left( {3\over 5},{4\over 5},0\right), \;
\left( 0,{4\over 5},{3\over 5}\right)
\in S^2\cap {\Bbb Q}^3.$$
The cross product thereof is
$$\left( {12\over 25},{-9\over 25},{12\over 25}\right)
\not\in S^2\cap {\Bbb Q}^3.$$

Indeed, if instead of  $ S^2\cap {\Bbb Q}^3$ one would start with
three non-orthogonal, non-collinear rational rays and generate
new ones by the cross product, one would end up with {\em all} rational rays
\cite{havlicek}.

In the Birkhoff-von Neumann approach to quantum logics \cite{birkhoff-36},
this non-closedness under
elementary operations such as the {\tt nor}-operation might be
considered a serious deficiency which rules out the above model as an
alternative candidate for Hilbert space quantum mechanics.
(However, his argument does not apply to the Kochen-Specker partial algebra
approach to quantum logic,
since there operations among propositions are only allowed if the
propositions are comeasurable.)

Informally speaking, the relative (with respect to other sets such as the
rational rays) ``thinness'' guarantees colorability.
In such a case, the formation
of finite cycles such as the ones introduced
by Kochen and Specker \cite{kochen1} are impossible.

To put it differently: given any nonzero measurement uncertainty
$\varepsilon$ and any
non-colorable Kochen-Specker graph $\Gamma (0)$
\cite{kochen1,svozil-ql},  there
exists another graph (in fact, a denumerable infinity thereof)
$\Gamma (\delta)$
which lies inside the range of measurement uncertainty
$\delta \le \varepsilon$
[and thus cannot be discriminated from the non-colorable
$\Gamma (0)$]
which {\it can} be colored.
Such a graph, however, might not be connected in the sense that the
associated subspaces can be cyclically rotated into itself by local
transformation along single axes. The set $\Gamma (\delta)$ might thus
correspond to a collection of tripods such that none of the axes
coincides with any other, although all of those non-identical single axes
are located within
$\delta$ apart from each other.
Indeed, this  appears to be precisely how the Clifton-Kent
construction works  \cite[p. 2104]{clifton:99}.

The reason why a Kochen-Specker type contradiction
does not occur in such a scenario is the impossibility
to ``close'' the argument; to complete the
cycle: the necessary propositions are simply not available
in the rational sphere model. For the same reason,
an equilateral triangle  does not exist in ${\Bbb Q}^2$ \cite{Specker-priv}.
Yet, while these findings seem to contradict the conclusions of Clifton and Kent \cite{clifton:99},
we would like to emphasize that this does not relate to their formal arguments but rather
is a matter of interpretation and a question of how much
should be sacrificed for value definiteness.

Thus, although the colorings of rational spheres offer a rather unexpected
possibility to define consistent classical models,
a closer examination shows
that any such colorings should be excluded for physical reasons.

\section*{Acknowledgments}
The authors would like to acknowledge stimulating discussions with Ernst Specker.

\end{document}